\documentclass[12pt,epsf]{article}

\usepackage{graphicx}
\def\){\right)}
\def\({\left( }
\def\]{\right] }
\def\[{\left[ }

\def\no{\nonumber \\}

\def\be{\begin{equation}}
\def\ee{\end{equation}}
\def\ba{\begin{eqnarray}}
\def\ea{\end{eqnarray}}
\def\no{\nonumber \\}

\begin{document}
\begin{titlepage}


\begin{center}
{\Large\bf Behavior of tachyon in string cosmology based on gauged WZW model}
\vspace{.4in}

{$\rm{Sunggeun \,\, Lee}^{\dag}$}\footnote{\it email:sglkorea@hotmail.com}\,\,and
{$\rm{Soonkeon \,\, Nam}^{\P}$}\footnote{\it email:nam@khu.ac.kr}\\

  {\small \dag \it Center for Quantum Spacetime, Sogang University, Seoul 121-742, Korea}\\
  {\small \P \it Department of Physics and Research Institute for Basic Sciences,
 Kyung Hee University, Seoul 130-701, Korea}\\

\vspace{.5in}
\end{center}
\begin{center}
{\large\bf Abstract}
\end{center}




\begin{abstract}
We investigate a string theoretic cosmological model in the context of the gauged Wess-Zumino-Witten model.
Our model is based on a product of non-compact coset space and a spectator flat space; $[SL(2,R)/U(1)]_k \times {\bf R}^2$. We extend the formerly
studied semiclassical consideration with infinite Kac-Moody level $k$ to a finite
one. In this case, the tachyon field appears in the effective action,
and we solve the Einstein equation
to determine the behavior of tachyon as a function of time. We find that tachyon field dominates over
dilaton field in early times.  In particular, we consider the energy
conditions of the matter fields consisting of the dilaton and the tachyon which affect the initial
singularity. We find that not only the strong energy but also the null energy condition is
violated.

\end{abstract}
\end{titlepage}

\section{Introduction}

As a strong candidate for a theory of quantum gravity, string theory should explain
our universe near the big bang, and should resolve the initial
singularity of the standard cosmology \cite{hawkpenr}.
Although a lot of effort has been made
to get reasonable answers, there are still more questions than answers. One of
the challenges of string cosmology \cite{gasper,quevedo,veneziano} is that, in general, string theory
is quite useful for static problems but for time dependent cases it
becomes much harder. For instance, string theory is quite
useful for resolutions of orbifold \cite{dixoharvvafawitt1,dixoharvvafawitt2} or conifold singularities \cite{strom},
or for explaining black hole entropy for extremal black holes \cite{stromvafa}, but it has not been
as successful in explaining black hole singularity \footnote{There has been
some attempts in this direction. For example, S. Mathur \cite{mathur} introduced the fuzzball
picture to get smooth geometry of the black hole.} itself or time dependent cosmological models \cite{craps,cornalbacosta}.

Nevertheless, the cosmological model in string theory can be studied in various ways.
One of them is to solve the Einstein equation from the low energy effective action
which comes from the vanishing beta function for conformal invariance of two dimensional string worldsheet action.
Another one, which is more stringy, is to study Wess-Zumino-Witten (WZW) action \cite{witten0,gepwit}
on a coset manifold  which is a gauged version of WZW model\cite{bardrabisaer,karaschn}.
In the approach using the effective string action, it is hard to see all the stringy effects.
It does not capture $\alpha'$ corrections to all orders.
In fact, solving the Einstein equations
with all higher $\alpha'$ corrections is very hard
except in theories with enough SUSY to cancel the higher derivative terms (when the non-renormalization theorem
kicks in).
On the other hand, the method based on the gauged WZW model captures physics to all orders of $\alpha'$ naturally.

One exact model for strings on
a curved spacetime was a model by Witten which is a gauged WZW model on $SL(2,R)/U(1)$ coset manifold \cite{witten}.
The target space of this background is a two dimensional black hole.
This model was further studied by Dijkgraaf, Verlinde, and Verlinde \cite{dijkgraafvv}
for the case of finite $k$.
Moreover, it can be extended to higher dimensions \cite{barssfet}.
Utilizing this solution, a four dimensional string theoretic cosmological model was also studied by taking
products of two cosets
$[SL(2,R)/U(1)]_{k_1}\times [SU(2)/U(1)]_{k_2}$ \cite{nappiwitten}. A very interesting cosmological
model based on gauged WZW model was introduced by Kounnas and L\"ust \cite{kounnaslust} who studied
$[SL(2,R)/U(1)]_k\times {\bf R}^2$, and by taking the level $k$ negative.
In this model, the singularity is absent, because it is hidden behind the horizon.

As mentioned above, the advantage of studying the WZW model is that
it provides an exact solution to
all orders in $\alpha'$. This model has a correction
through the Kac-Moody level $k$ by $1/k$. Considering the leading
order term in $1/k$ expansion is equivalent to considering the problem in semiclassical limit.
With the stringy background obtained from the gauged WZW model in leading order $1/k$,
Kounnas and L\"ust studied the effect of the dilaton field using the effective action.
This means that, while the usual string effective action
should be checked to all higher orders in $\alpha'$ for requiring exact CFT (vanishing beta function),
the WZW action is correct for all $\alpha'$, the only correction is in terms of level $k$.

In this paper we extend the work of Kounnas and L\"ust to all orders in $1/k$. We find
the cosmological metric for finite $k$ by taking $k$ negative of the metric
in Refs. \cite{dijkgraafvv,sfet0}. The gauged WZW model provides an exact stringy background,
and we will study the effects of fields in the background. We include extra two dimension as was done in
Ref. \cite{kounnaslust} to
study as a four dimensional cosmology. After setting up the model, we solve the cosmological
model in the four dimensional
effective action including the tachyon as well as the dilaton. Especially,
the time dependence of the tachyon is considered.

The rest of the paper is organized as follows:
In section II we review the model by Kounnas and L\"ust which is the case of semiclassical consideration.
In section III we discuss the case of finite $k$, and also consider the effect of tachyon in the
background. We study various energy conditions applied to the matter
fields (dilaton and tachyon) in this case. In section IV we conclude with discussions.

\section{Kounnas and L\"ust model: $k\to \infty$ case}

In this section, we briefly review the work of Kounnas and L\"ust\cite{kounnaslust}.
They considered a gauged WZW model based on the coset $[SL(2,R)/U(1)]_k\times {\bf R}^2$.
The gauged WZW model on the coset $G/H$ is basically described by the following action \cite{bardrabisaer,karaschn}:
\ba
S&=&{k\over {4\pi}} \int d^2 z (g^{-1}\partial g g^{-1}\bar \partial g)-{k\over {12\pi}} \int_B
tr(g^{-1}dg\wedge g^{-1}dg\wedge g^{-1}dg) \no
&+&{k\over {2\pi}}\int d^2z tr(A\bar \partial g g^{-1}-\bar A g^{-1}\partial g -g^{-1}Ag \bar A^{-1}).
\ea
Here, the boundary of $B$ is the 2 dimensional worldsheet, $g$ is a group element of the group $G$, and
$A$ is the gauge field of $H$, which is a subgroup of $G$. The level of the Kac-Moody algebra
for the group of $G$ is $k$. Here $\partial$ and $\bar \partial$ denote
partial integration with respect to the complex world sheet coordinates $z$ and $\bar z$, respectively.

The central charge of the WZW model on the compact $SU(2)/U(1)$ coset is $c=3k/(k+2)-1$, while for noncompact
coset $SL(2,R)/U(1)$ it is $c=3k/(k-2)-1$.
For the WZW model on the coset $[SL(2,R)/U(1)]\times {\bf R}^2$, we get central charge $c=3k/(k-2)-1+2=4+6/(k-2)$.

We can parametrize $G=SL(2,R)$ as
\be
g=\left(
\begin{array}{cc}
u&a \\
-b&v \\
\end{array}
\right)
\ee
Choosing a gauge with $a=\pm b$, the action can be written as a sigma model action of the form
\be
S=\int d^2z g_{\mu\nu}(X)\partial X^\mu \bar \partial X^\nu .
\ee
That is, the gauged WZW model describes strings moving on some coset manifold with
the target space described by coordinate $X^\mu$ with spacetime metric $g_{\mu\nu}(X)$.
After fixing the gauge, we see that the
WZW model is reduced to a non-linear sigma model.
For example, for the coset $SL(2,R)/U(1)$, the target space is a black hole
described by a metric \cite{witten}
\be
ds^2=-k{ dudv\over {1-uv}}.
\ee

If we are interested in four dimensional target spacetime, the simplest case is when we include two extra
flat dimensions ${\bf R}^2$. The coset is now $[SL(2,R)/U(1)]\times {\bf R}^2$ and for the semiclassical limit
$k\to \infty$, the four dimensional sigma model metric is
\be
ds_\sigma^2=-k {dudv\over {1-uv}}+dx_2^2+dx_3^2,
\ee
where $x_2$ and $x_3$ are the coordinates of ${\bf R}^2$.
So the metric describes a two dimensional black hole background with two spectator dimensions.
If we introduce coordinates $x_0$ and $x_1$ such that
\be
u=e^{x_1}\sinh x_0,~~~{\rm and}~~~v=-e^{-x_1}\sinh x_0,
\ee
we get another form of the metric:
\be
ds^2=-k(-dx_0^2 +\tanh^2 x_0 dx_1^2 ) +dx_2^2+dx_3^2.
\ee
Near $x_0\to 0$, the dilaton vanishes and the metric becomes completely smooth, being flat, when $x_0$ is non-compact.
The spacetime is a product of two dimensional Milne universe times ${\bf R}^2$ \cite{craps0}. Although $t=0$ is non-singular, this
is as close as we can get to the big bang singularity in this model.
In order to have a time dependent cosmological background, we need to
make $x_0$ coordinate timelike. So we will denote $x_0$ as $t$.
This is achieved by making the level $k$ negative. As we can see from the scale
factor, $\tanh^2 t$, this
cosmological metric is not good for realistic cosmology such as for inflationary.
The scale factor saturates at late times. However, the metric
could describe the spacetime near the big bang. Since our starting point is a gauged WZW model which
has stringy nature in itself, we might expect to have a resolution of the singularity.
Indeed, in the model of Kounnas and L\"ust, by taking negative $k$, they
could obtain singularity free spacetime background. Choosing negative
$k$ has an effect of rotating the Penrose diagram of the black hole.
As a result, the big bang singularity is hidden behind what
used to be the black hole horizon. Hence, from now on, we take $k$ as negative and rename $-k$ as $k$.
Taking negative Kac-Moody level may give rise to non-unitarity but it is not solved completely
\cite{dixpeslyk,hwang}. We ignored this issue in the present work.

By requiring conformal invariance of the non-linear sigma model, i.e., vanishing of the beta function, we get
the four dimensional effective action for the graviton $g_{\mu\nu}$ and the dilaton $\phi$ background. In
string frame it is given by the following action \cite{callmartperrshen}:
\be
S=\int d^4 x \sqrt {-det(g_{\mu\nu})} e^{-\phi} \( R+(\nabla \phi)^2 + \Lambda\),~~~~~\Lambda =-4/(k+2),
\ee
Here, $\Lambda$ which plays the role of cosmological constant,  and we have put $8\pi G=1$, where $G$ is the Newton's constant.
The dilaton is given by
\be
\phi=-\log(1-uv)=-\log(1+\sinh^2 t)=-\log(\cosh^2 t),\label{dilaton}
\ee
such that $e^\phi=1/\cosh^2 t$.

The Einstein equations which follow from the action are:
\ba
&&R_{\mu\nu} + \nabla_\mu\nabla_\nu \phi=0 , \no
&&R-(\nabla_\mu \phi)^2 +2\nabla_\mu\nabla^\mu \phi+\Lambda=0.
\ea

Now let us consider the equation in the Einstein frame.
We can get the Einstein frame action by conformal transformation $g_{E,\mu\nu}=e^{-\phi}g_{Str,\mu\nu}$.
As a result we get the Einstein frame action
\be
S=\int d^4 x \sqrt{-det(g_{\mu\nu})} \(R-{1\over 2}(\nabla \phi)^2 -e^\phi {4 \over {k+2}}\).
\ee
The Einstein equation is
\be
R_{\mu\nu}-{1\over 2}g_{\mu\nu}R=T^\phi_{\mu\nu},
\ee
where
\be
T^\phi_{\mu\nu}= {1\over 2}\nabla_\mu\phi \nabla_\nu \phi -{1\over 4}g_{\mu\nu}(\nabla \phi)^2
-{2 \over { k+2}} e^\phi g_{\mu\nu}.
\ee
In the Einstein frame, the metric is now given by
\be
ds^2=\cosh^2 t \[ (k+2)(-dt^2+\tanh^2tdx_1^2)+dx_2^2+dx_3^2 \].
\ee

We find that this metric is non-singular.
In order to check this, we calculate the scalar curvature which is given as follows:
\be
R=-{ {(\sinh^2 t +2)}\over {2(\sinh^2 t +1)}}.
\ee
This metric is obviously nonsingular as $t\to 0$. As we expected, string theory prevents initial big bang
singularity in this model.

Now let us see the behavior of the dilaton field. It is quite illuminating to check
the energy conditions. There is a theorem by Hawking and Penrose \cite{hawkpenr} which
states that there must be a singularity when the strong energy
condition, $\rho+3p\ge 0$ is satisfied. Here, we expect to find the
violation of this energy condition. Now, let us calculate the components of the stress
energy tensor $T_{00}$ and $T_{ii}$ to check energy conditions.
\be
T^\phi_{00}={1\over 2}\nabla_0 \phi \nabla_0 \phi -{1\over 4} g_{00} \nabla_0 \phi \nabla_0 \phi g^{00}
-{2\over {k+2}}g_{00}e^{\phi}={ {3\tanh^2t+2} \over {(1+\tanh^2t)^2}}=\rho,
\ee
and
\ba
&&T^\phi_{ii}={1\over 2} \nabla_i \phi \nabla_i \phi -{1\over 4} g_{ii} (\nabla \phi)^2
-{2 \over {k+2}}e^\phi g_{ii}=g_{ii}\(-{1\over 4} g^{00} \dot \phi^2 -{2 \over {k+2}}e^\phi\)\no
&&~~~=g_{ii} {1\over {k+2}}{{-\tanh^2t -2} \over { (1+\tanh^2t)^2}}=pg_{ii}.
\ea
We can absorb the $1/(k+2)$ into the metric such that
$T^\phi_{ii}=g_{ii}(-\tanh^2t -2) /  (1+\tanh^2t)^2$.

For our further consideration of energy conditions we summarize the
four energy conditions, null (NEC), weak (WEC), strong (SEC), and dominant (DEC) energy conditions:
\ba
&&{\rm NEC}:~\rho+p \geq 0,\no
&&{\rm WEC}:~\rho\geq 0 ~~{\rm and} ~~\rho+p \geq 0,\no
&&{\rm SEC}:~\rho+3p \geq 0 ~~{\rm and} ~~\rho+p\geq 0,\no
&&{\rm DEC}:~\rho\geq 0 ~~{\rm and}~~\rho\pm p\geq 0.
\ea

If we check the weak energy condition and strong energy condition for Kounnas and L\"ust model,
we have the following:
\ba
&&{\rm WEC}: ~\rho+p={2 \tanh^2t \over {(1+\tanh^2t)^2}}>0, \no
&&{\rm SEC}:~\rho+3p=-{4\over {(1+\tanh^2t)^2}} <0.
\ea
We see that only the strong energy condition is violated.
Below we draw a picture for both energy conditions.

\begin{figure}
\begin{center}
\includegraphics[width=2.4in]{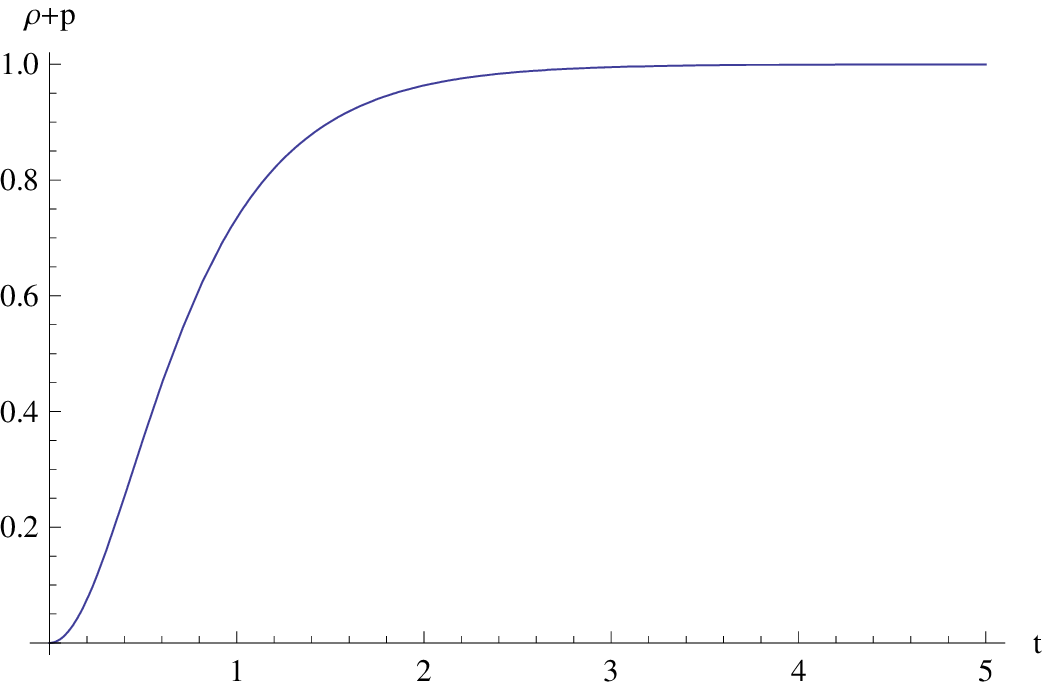}
\includegraphics[width=2.4in]{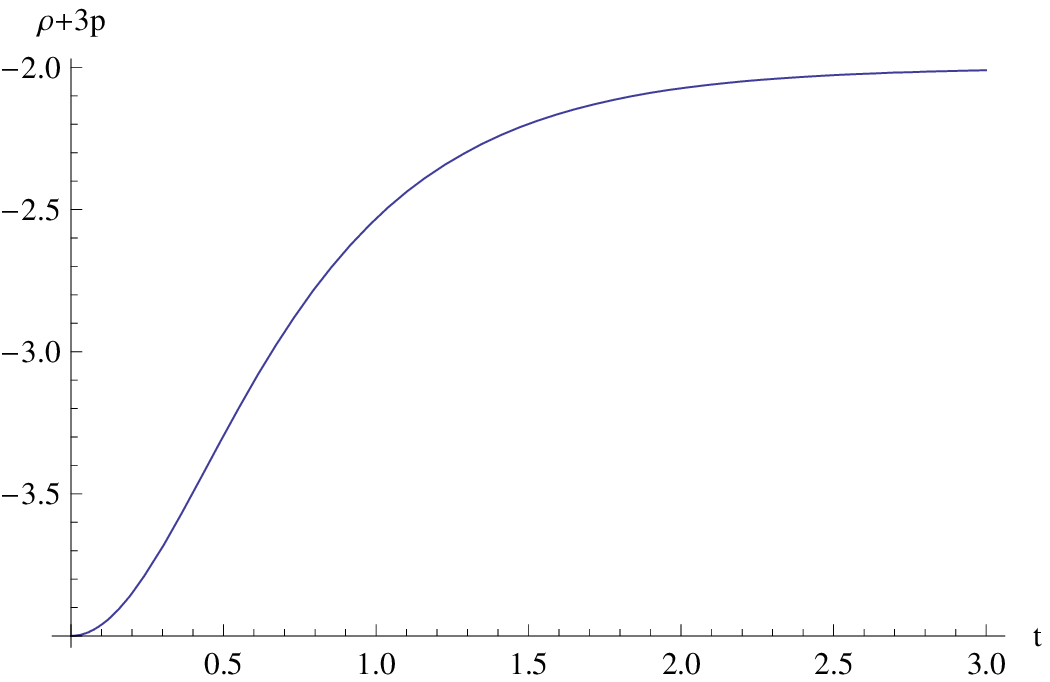}
\end{center}
\caption{The behavior of $\rho+p$ and $\rho +3p$. We see that $\rho+3p$ is
definitely negative and violates weak energy condition.}
\end{figure}

The violation of the strong energy condition is consistent with
the absence of the big bang singularity of the background geometry.

\section{The behavior of tachyon field: finite level $k$ }

We now consider the same coset model considered by Kounnas and L\"ust, but with the metric corrected by finite value of $k$.
The metric for the $[SL(2,R)/U(1)]_k$ WZW model with finite $k$ was studied in Refs. \cite{dijkgraafvv,sfet0}.
Again we include extra two directions $x_2$ and $x_3$ such that we have $[SL(2,R)/U(1)]\times {\bf R}^2$ coset space to study
four dimensional spacetime but they do not play much role.
To get a cosmological metric we take $k$ negative. As a result we get the the metric in Einstein
frame which is given by the following:
\be
ds^2 = \( \cosh^2 t \sqrt{ 1+{2\over k} \tanh^2 t} \) \[ { {k+2 }\over 2} \(-dt^2 +
{1\over {\coth^2 t +{2\over k}}}dx_1^2 \) +dx_2^2 +dx_3^2\].
\ee
The dilaton $\phi$ is given by
\be
e^{-\phi} =\cosh^2 t \sqrt{1+{2\over k}\tanh^2 t}\label{dilaton2}.
\ee
As $k\to \infty$, $e^\phi$ reduces to Eq.(\ref{dilaton}).
As in the case of semiclassical approach $k\to \infty$, the metric is obviously singularity free.
In fact, we have checked that the metric has
no singularity, with the help of ${\rm Mathematica}^{\rm TM}$. The expression is
to complicated and we will not write it down here. The string coupling constant, $g_s \sim e^\phi$,
becomes weak as time flows to future with initially finite value.

Now that the spacetime action contains the tachyon ${\cal T}$ we would like to see the
time dependent behavior of tachyon field.
The action in Einstein frame with the tachyon is given by \cite{tsey,ghoro}
\be
S=\int d^4 x \sqrt{-det(g_{\mu\nu})} \(R-{1\over 2} (\nabla \phi)^2 +\lambda e^\phi +(\nabla {\cal T})^2 -V({\cal T})e^\phi \),~~~
\lambda=-{4\over {k+2}}.
\ee
Here $V({\cal T})$ is a tachyon potential whose exact form is not necessary.
It is hard to solve the Einstein equations of motion from
this action. However, since the metric was already determined by fixing the coset
manifold, we just need to find
the tachyon field as a function of time by solving the Einstein equations of motion of this action.
For this we will assume that all the fields including the tachyon field ${\cal T}$ depends only on the time $t$.

By varying the above action, we get field equations including the Einstein equations, which are given by
\ba
&&R_{\mu\nu} -{1\over 2}g_{\mu\nu} R \no
&&~~~~~= {1\over 2}\nabla_\mu \phi \nabla_\nu \phi
-{1\over 4} g_{\mu\nu} (\nabla\phi)^2 +{\lambda\over 2} g_{\mu\nu}e^\phi
-{1\over 2}g_{\mu\nu}e^\phi V({\cal T})-\nabla_\mu {\cal T} \nabla_\nu {\cal T} +{1\over 2}g_{\mu\nu}(\nabla {\cal T})^2 \no
&&\ \ \ \ \ \  =T^\phi_{\mu\nu}+T^{\cal T}_{\mu\nu}=T_{\mu\nu}, \no
&&\nabla^2 \phi +\(\lambda -V({\cal T})\) e^\phi =0, \no
&&-2\nabla^2 {\cal T} - {d \over {d{\cal T}}} V({\cal T}) e^\phi =0.\label{eqm01}
\ea

To find the time dependent behavior of the tachyon ${\cal T}$, we put the solutions for metric and
the dilaton into the above equations. However, it is still quite hard to solve them analytically.
So, we take approximations and see the asymptotic
behavior of the tachyon field. Another thing to check is that
since the metric has no singularity, we have to check to see whether the weak- and strong-energy conditions
violate or not.

The energy momentum tensor can be read from above after some calculation.
With the help of the equations of motions (\ref{eqm01}), $T_{00}$ and $T_{ii}$ can be reduced to the following:
\ba
&&T_{00}={1\over 4}{\dot \phi}^2 -{1\over 2}g_{00}\nabla^2 \phi -{1\over 2}{\dot {\cal T}}^2, \no
&&T_{ii}= -{1\over 4}g_{ii} g^{00} {\dot \phi}^2 -{1\over 2} g_{ii}\nabla^2 \phi +{1\over 2}g_{ii}g^{00} {\dot {\cal T}}^2,
\ea
where we have denoted the time derivative $d\over dt$ by $\cdot  $.
Moreover, due to the explicit form of the dilaton $\phi$ in Eq.(\ref{dilaton2}), we just need to know the $\dot {\cal T}$
in order to determine $T_{00}$ and $T_{ii}$.

Since we can put
\be
{d \over {d{\cal T}}} V =\dot V /\dot {\cal T},~~~ {\rm and} ~~~\dot V ={d \over {dt}}(e^{-\phi}\nabla^2 \phi)
=-\dot \phi e^{-\phi} \nabla^2 \phi
+ e^{-\phi} {d \over {dt}} (\nabla^2 \phi),
\ee
we need to solve the following equation to get the equation for $\dot {\cal T}$:
\be
-2\dot {\cal T} ={  \dot V e^\phi \over \nabla^2 {\cal T}} \ \ \ \ \ \ \leftrightarrow \ \ \ \ \ \ \dot {\cal T} e^{-\phi}\nabla^2 {\cal T} =-{1\over 2} \dot V
=-{1\over 2}{d\over {dt}} (e^{-\phi}\nabla^2 \phi).
\ee
In other words, the tachyon ${\cal T}$ satisfies the following equation:
\be
\dot {\cal T} e^{-\phi}\nabla^2 {\cal T} =-{1\over 2}{d\over {dt}} (e^{-\phi}\nabla^2 \phi),
\ee
which reads
\be
\ddot {\cal T} + {\dot f \over f } \dot {\cal T} =-{1\over 2}{1\over {\dot {\cal T}}} {d \over {dt}} \left(\ddot \phi
+ { \dot f \over f}\dot \phi\right).
\ee
In the above we have introduced
\ba
&&f(t)
=g^{00}\sqrt{-\det(g)}
=-\sinh t \cosh t.
\ea

The time dependence of the tachyon potential, $V(t)$, satisfies the following differential equation:
\be
\dot V={d \over {dt}} \[ -{2\over {k+2}} {1\over {fg}} g\(\dot f \dot \phi +f\ddot \phi\)\]=-{2 \over {k+2}}
{d\over {dt}}\( \ddot\phi+{\dot f \over f} \dot \phi \).
\ee

In terms of known functions $f(t)$ and $\phi(t)$ in Eqs.(\ref{dilaton}) and (\ref{eqm01}), we can rewrite the $T_{00}$ and $T_{ii}$ as
\ba
T_{00}&=&{1\over 4}{\dot\phi}^2 -{1\over 2} (\ddot \phi +{\dot f \over f}\dot \phi) -{1\over 2}{\dot {\cal T}}^2,
\no
T_{ii}&=&-{1\over 4} g_{ii}g^{00}{\dot \phi}^2 +{1\over 2} g_{ii} (-\nabla^2 \phi)+{1\over 2}g_{ii}g^{00}{\dot {\cal T}}^2\no
&=&g_{ii}\left(-{1\over 4} g^{00}{\dot \phi}^2 -{1\over 2}\nabla^2 \phi +{1\over 2}g^{00}{\dot {\cal T}}^2\right)\no
&=&g_{ii} e^\phi \({1\over 4} {\dot \phi}^2 +{1\over 2} (\ddot \phi +{\dot f \over f}\phi)\)
-g_{ii} e^\phi {1\over 2}{\dot {\cal T}}^2,
\ea
where we absorbed ${2\over {k+2}}$ into the metric, and used the following relation:
\be
g_{00}\nabla^2 \phi =\ddot \phi+{ \dot f \over f }\dot \phi.
\ee

For later convenience, we calculate the following expressions:
\ba
&&(\dot \phi)^2={ {4\tanh^2 t (2+k-\cosh^{-2} t)^2 }\over { (k+2\tanh^2 t)^2}},\no
&&{\dot f \over f}=\coth t+\tanh t , \no
&&{d\over {dt}}\left(\ddot \phi + {\dot f\over f}\dot \phi\right)=F(t),
\ea
where
\be
F(t)\equiv -{{ (12+4k+3k^2+4(k^2-4)\cosh 2t +(2+k)^2\cosh 4t )\cosh^{-4}t}
\over {2(k+2\tanh^2 t)^2}}.
\ee
Now, $T_{00}$ and $T_{ii}$ can be expressed as follows:
\ba
&&T_{00}=\rho= \rho^--{1\over 2}{\dot {\cal T}}^2 ,\no
&&T_{ii}=g_{ii}p=g_{ii}p^--g_{ii}e^\phi {1\over 2}{\dot {\cal T}}^2.
\ea
where
\ba
&&\rho^- \equiv { {3(2+k)^2-(2+k)(12+k)\cosh^{-2}t +(13+4k)\cosh^{-4}t-\cosh^{-6}t}
\over {(k+2\tanh^2 t)^2}},\no
&&p^- \equiv {-{ \cosh^{-2}t((2+k)^2 +(k-4)(2+k)\cosh^{-2}t+3\cosh^{-4}t +\cosh^{-6}t)}
\over {(k+2\tanh^2 t)^2 (1+{2\over k}\tanh^2 t)^{1\over 2}}}.
\ea

Now let us consider solving the differential equations for tachyon and
obtain the behavior of the tachyon field. We have to solve the following equation:
\be
\ddot {\cal T} + {\dot f \over f } \dot {\cal T} =-{1 \over {\dot {\cal T}}} {d \over {dt}} \left(\ddot \phi + { \dot f \over f}\dot \phi \right),
\ee
which can also be rewritten as
\be
\ddot {\cal T}+ (\coth t +\tanh t) \dot {\cal T}=-{1\over 2}{1\over {\dot {\cal T}}}F(t).
\ee
This differential equation is the Bernoulli differential equation:
\be
\dot {\cal G} -P(t) {\cal G}=Q(t) {\cal G}^n
\ee
where for ${\cal G}={\dot {\cal T}}$ the solution is given by
\be
{\cal G}^{1-n}=ce^F+(1-n)e^F \int e^{-F}Q(t) dt,
\ee
where
\be
F=(1-n)\int P(t) dt.
\ee

This equation is still hard to solve analytically in our case. Hence, we take two opposite limits to see
the asymptotic behavior.
In the limit, $t\to \infty$, and $t\to 0$, this equation simplifies a lot.
\ba
&& t\to \infty:~~\ddot {\cal T} +2\dot {\cal T} ={2 \over {\dot {\cal T}}}, \no
&&t\to 0 :~~\ddot {\cal T} +t^{-1} \dot {\cal T} ={{2(1+k)}\over k}{1\over \dot {\cal T}}.
\ea
The first equation has a solution
\be
{\dot {\cal T}}^2 =1+e^{-4(t-t_0)},
\ee
and the solution for the second equation is
\be
{\dot {\cal T}}^2 = {c\over t^2} +{4\over 3}{{k+1}\over k}t,
\ee
where $t_0$ and $c$ are integration constants. When we assume the tachyon field
is a real value then $c$ is a positive constant.

We note that the tachyons becomes negligible in later times. The important contribution
of the tachyons are in the early time near big bang ($t \sim 0$). In fact, it dominates over other fields for
the contributions to $\rho$ and $p$. Due to tachyons, both $\rho$ and $p$ become negative values.
Hence, we see that even the null energy condition is violated.

With the given energy momentum tensor we see that the
conditions are calculated as follows for both null $p+\rho$ and strong $3p+\rho$:
\ba
&&{\rm NEC}:~ \rho+p=\rho^-+p^--{1\over 2}{\dot {\cal T}}^2 \(1+\cosh^{-2}t(1+{2\over k}\tanh^2 t)^{-{1\over 2}}\),\no
&&{\rm SEC}:~ \rho+3p=\rho^-+3p^--{1\over 2}{\dot {\cal T}}^2 \(1+3\cosh^{-2}t(1+{2\over k}\tanh^2 t)^{-{1\over 2}}\).
\ea

\begin{figure}[hbt]
\begin{center}
\includegraphics[width=2.4in]{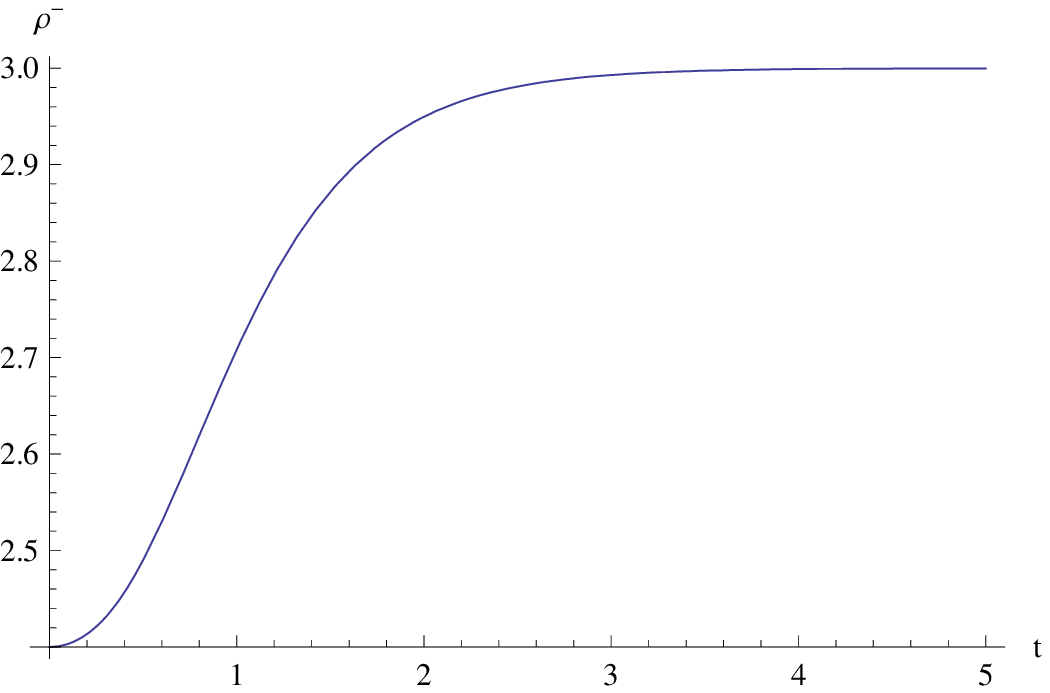}
\includegraphics[width=2.4in]{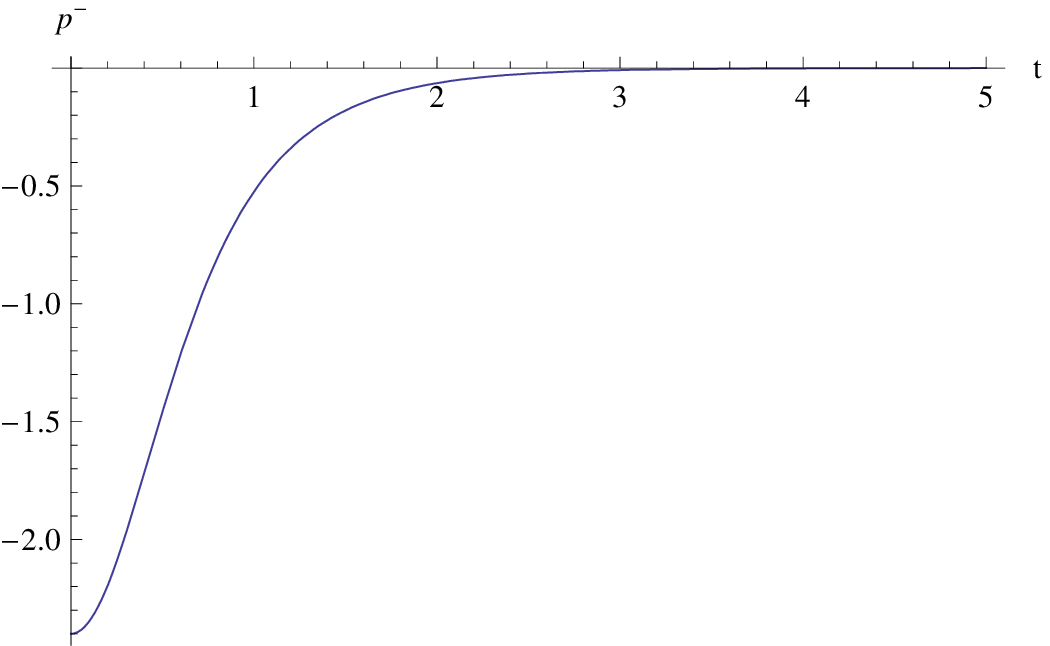}\\
\includegraphics[width=2.4in]{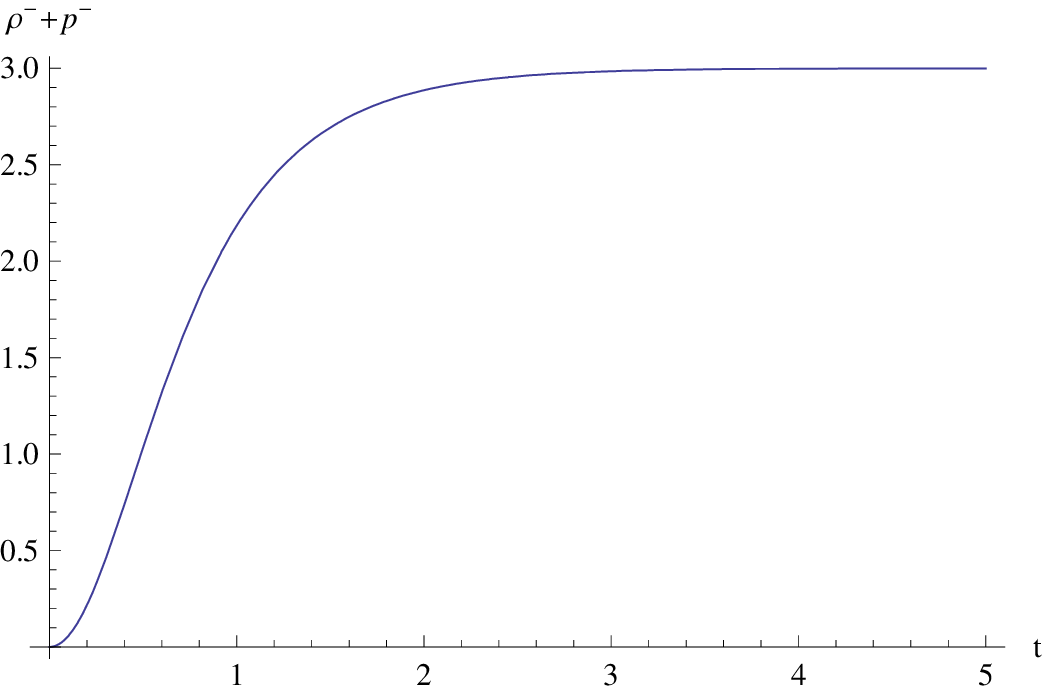}
\includegraphics[width=2.4in]{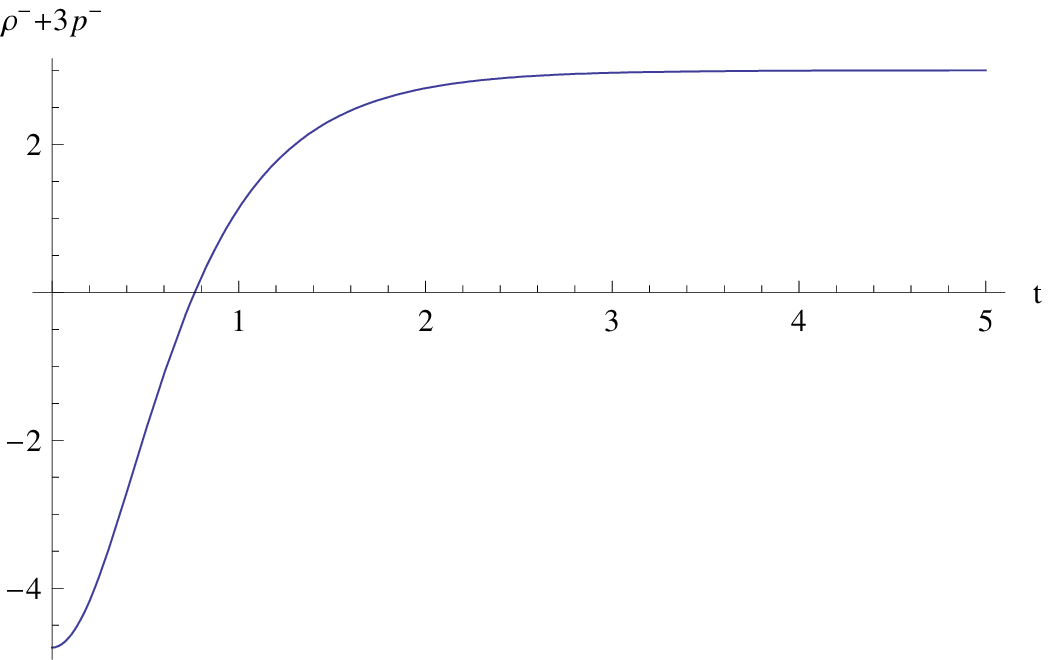}
\end{center}
\caption{The behavior of $\rho^-$, $p^-$, $\rho^-+p^-$ and $\rho^- +3p^-$ as a function of time. In these
diagrams the kinetic term of the tachyon is not involved but the contribution of the tachyon. The energy density
is definitely negative.}
\end{figure}

In figure 2, we plot the $\rho$, $p$, $\rho +p$ and $\rho+3p$ below. The kinetic term
of the tachyon, $\sim \dot {\cal T}^2$, is not involved temporarily. The only effect of tachyon comes from the potential.
The tachyonic contribution is dominant near the big bang $t\sim 0$ and is negligible in later time.
Here we have violation of the null energy condition as well as the violation
of the strong energy condition. This is again consistent with
the absence of the initial big bang singularity of the background geometry. In Ref.\cite{brustmadde},
the authors discussed that for the graceful exit the matter that violates the null energy condition
is necessary. We would like to
comment on the violation of the null energy condition.
It seems that the violation of the null energy condition when the tachyon
dominates does not seem to be problematic. Tachyon is just a signal of
instability. We can interpret this tachyon as a supplementary matter for
wormhole \cite{morrthoryurt,misnerwheeler,wheeler} which mediates the topology change near the big bang.
It is believed that the wormhole ia a part of the space time foam bubbling near the big bang
which quantum mechanical picture is quite necessary.
So our model seems to provide useful ingredient for the early universe.

\begin{figure}
\begin{center}
\includegraphics[width=2.4in]{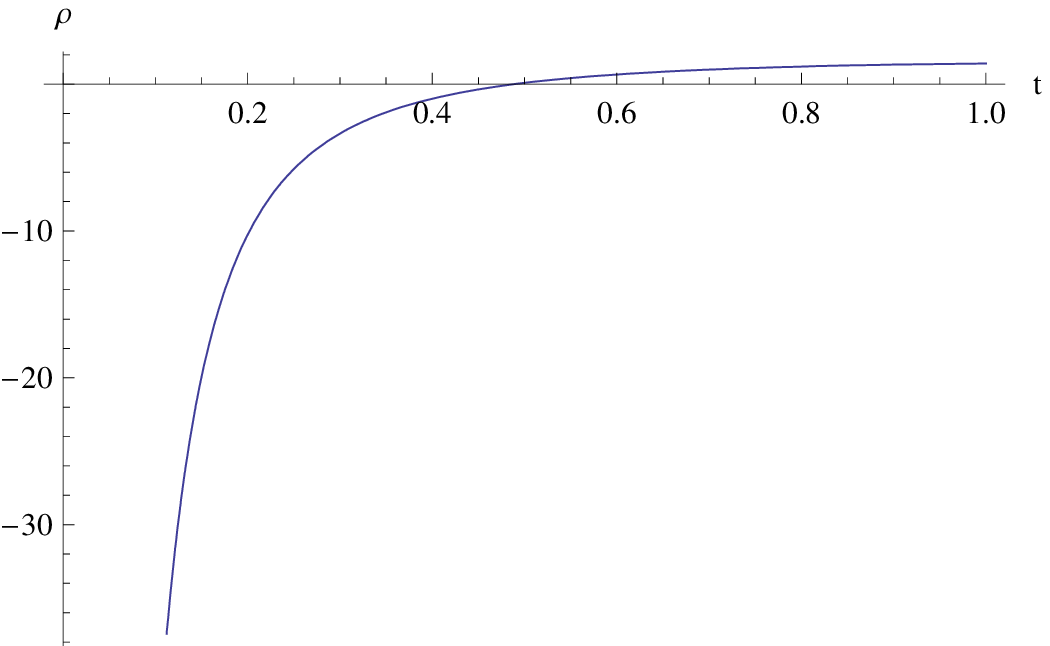}
\includegraphics[width=2.4in]{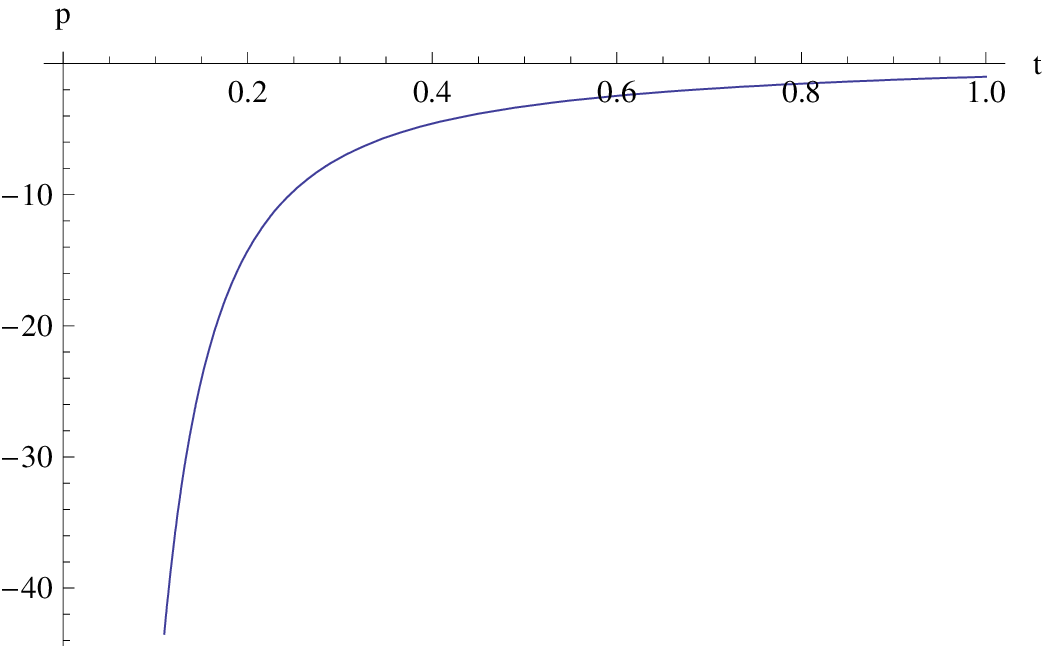}\\
\includegraphics[width=2.4in]{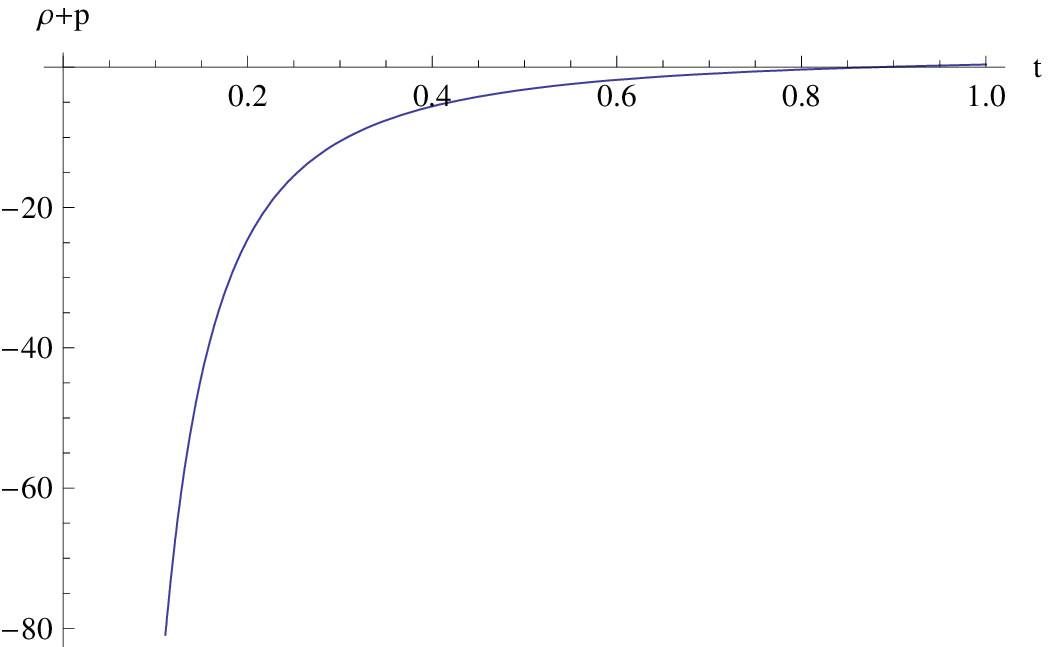}
\includegraphics[width=2.4in]{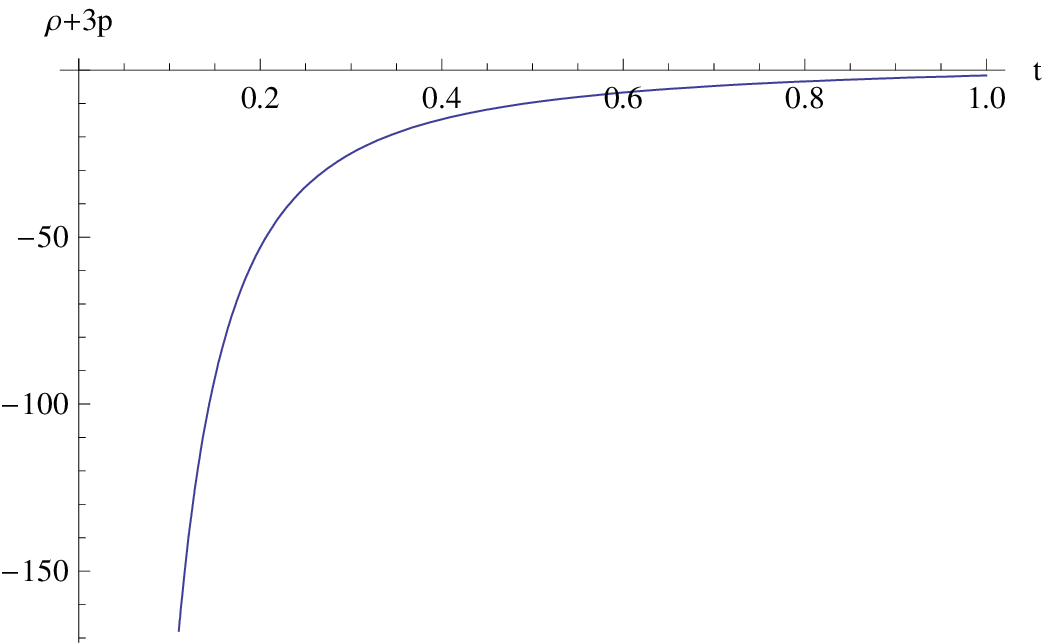}
\end{center}
\caption{The behavior of $\rho$, $p$, $\rho+p$ and $\rho +3p$ at $t\sim 0$ with the inclusion of tachyon
kinetic term. Note that the energy density itself is negative due to
tachyon effect. In addition, both the strong and null energy condition
are violated.}
\end{figure}

In figure 3, we plot the $\rho$, $p$, $\rho +p$ and $\rho+3p$ again
including the tachyon kinetic contribution. Because of negligible tachyon effect
in later time, we only focus on the times near big bang $t\sim 0$. Here we notice that
in addition to the violation of the both strong and weak energy conditions,
the energy density itself becomes negative near the big bang where the tachyon
effect is formidably big.

\section{Conclusion }

In this paper, we studied the behavior of the tachyon in a cosmology based on the gauged WZW model.
The cosmological metric is obtained by taking the negative sign of the Kac-Moody level $k$.
We used the smooth, singularity free metric obtained in Refs.\cite{barssfet,dijkgraafvv,sfet0}.
Next, we considered the behavior of tachyon field in the metric and the dilaton background by solving the equation
of motion of the effective action. Although, the fully explicit solution could in principle be found by solving
the Bernoulli differential equaion, we found on the initial ($t\to 0$) and asymptotic ($t\to \infty$) bahevior of the tachyon field.
We find that tachyon field dominates in early tims. We could then calculate the components
of the energy momentum tensor, $T_{00}$ and $T_{ii}$, and checked the energy conditions in connection to the singularity theorem. Contrary to the result of
Kounnas and L\"ust\cite{kounnaslust} where only the strong energy condition was violated, we have the strong as well as the null energy conditions violated near
big bang ($t\sim 0$), where the tachyon dominates.
It seems that the tachyon domination in the early universe might
support understand the wormholes in the early universe, which is believed to be a space time foam.

Our consideration can be applied to other backgrounds such as
Nappi-Witten model \cite{nappiwitten}. Moreover, we can consider other coset models
or inclusion other fields such as antisymmetric tensor fields.
It is also interesting to see the topology changing studied in Ref.\cite{kiritkouna}

\section{Acknowledgements}
This work was supported by the Science Research Center Program of
the Korea Science and Engineering Foundation through
the Center for Quantum Spacetime(CQUeST) of
Sogang University with grant number R11 - 2005 - 021.
It was also supported by the Korea Research Foundation Grant funded by the
Korean Government(MOEHRD) (KRF-2007-314-C00056 ).


\begin{thebibliography}{0}
\bibitem{hawkpenr} S.~W.~Hawking and G.~F.~R.~Ellis, "The large scale of space-time,''
Cambridge Univ.~Press, Cambridge, 1973.
\bibitem{gasper} M.~Gasperini, ``Elements of string cosmology," Cambridge Univ. ~Press, Cambridge, 2007.
\bibitem{quevedo} F.~Quevedo,  Class.\ Quant.\ Grav.\  {\bf 19} (2002) 5721.
\bibitem{veneziano} G.~Veneziano,~hep-th/0002094 and references therein.
\bibitem{dixoharvvafawitt1} L.~J.~Dixon, J.~A.~Harvey, C.~Vafa and E.~Witten, Nucl.\ Phys.\  B {\bf 261} (1985) 678.
\bibitem{dixoharvvafawitt2} L.~J.~Dixon, J.~A.~Harvey, C.~Vafa and E.~Witten, Nucl.\ Phys.\  B {\bf 274} (1986) 285.
\bibitem{strom} A.~Strominger, Nucl.\ Phys.\  B {\bf 451} (1995) 96.
\bibitem{stromvafa} A.~Strominger and C.~Vafa, Phys.\ Lett.\  B {\bf 379} (1996) 99.
\bibitem{craps}  B.~Craps,  Class.\ Quant.\ Grav.\  {\bf 23} (2006) S849.
\bibitem{cornalbacosta} L.~Cornalba and M.~S.~Costa, Fortsch.\ Phys.\  {\bf 52} (2004) 145 and references therein.
\bibitem{mathur} S.~D.~Mathur, Fortsch.\ Phys.\  {\bf 53} (2005) 793.
\bibitem{witten0} E.~Witten, Commun.\ Math.\ Phys.\  {\bf 92} (1984) 455.
\bibitem{gepwit} D.~Gepner and E.~Witten, Nucl.\ Phys.\  B {\bf 278}, 493 (1986).
\bibitem{bardrabisaer} K.~Bardacki, E.~Rabinovici and B.~Saering, Nucl.~Phys.~B {\bf 301} (1988) 151.
\bibitem{karaschn} D.~Karabali and H.~J.~Schnitzer, Nucl.~Phys.~B {\bf 329} (1990) 649.
\bibitem{witten} E.~Witten, Physi.~Rev. {\bf D 44} (1991) 314.
\bibitem{barssfet} K.~Sfetsos, Phys.\ Rev.\  D {\bf 46} (1992) 4510.
\bibitem{nappiwitten} C.~R.~Nappi and E.~Witten, Phys.\ Lett.\  B {\bf 293} (1992) 309.
\bibitem{kounnaslust} C.~Kounnas and D.~L\"ust, Phys.\ Lett.\  B {\bf 289} (1992) 56.
\bibitem{dijkgraafvv} R.~Dijkgraaf, H.~L.~Verlinde and E.~P.~Verlinde, Nucl.\ Phys.\  B {\bf 371} (1992) 269.
\bibitem{sfet0} K.~Sfetsos, Nucl.\ Phys.\  B {\bf 389} (1993) 424.
\bibitem{callmartperrshen} C.~G.~Callan, E.~J.~Martinec, M.~J.~Perry and D.~Friedan,
Nucl.\ Phys.\  B {\bf 262} (1985) 593.
\bibitem{craps0}   B.~Craps, D.~Kutasov and G.~Rajesh, JHEP {\bf 0206} (2002) 053.
\bibitem{tsey} A.~A.~Tseytlin, Phys.\ Lett.\  B {\bf 264} (1991) 311.
\bibitem{ghoro} K.~Ghoroku, Phys.\ Lett.\  B {\bf 347} (1995) 21.
\bibitem{brustmadde} R.~Brustein and R.~Madden, Phys.\ Lett.\  B {\bf 410} (1997) 110.
\bibitem{morrthoryurt} M.~S.~Morris, K.~S.~Thorne and U.~Yurtsever, Phys.\ Rev.\ Lett.\  {\bf 61} (1988) 1446.
\bibitem{misnerwheeler}  C.~W.~Misner and J.~A.~Wheeler, Annals Phys.\  {\bf 2} (1957) 525.
\bibitem{wheeler} J.~A.~Wheeler, Annals Phys.\  {\bf 2} (1957) 604.
\bibitem{dixpeslyk} L.~J.~Dixon, M.~E.~Peskin and J.~D.~Lykken, Nucl.\ Phys.\  B {\bf 325} (1989) 329.
\bibitem{hwang} S.~Hwang, Nucl.\ Phys.\  B {\bf 354} (1991) 100.
\bibitem{kiritkouna} E.~Kiritsis and C.~Kounnas, Phys.\ Lett.\  B {\bf 331} (1994) 51.

\end{thebibliography}
\end{document}